\tolerance = 10000
\magnification = \magstep1


\def \Eff {{\bf F}}
\def \Que {{\bf Q}}
\def \Ree {{\bf R}}
\def \Zee {{\bf Z}}

\def \Fq  {{\Eff_q}}
\def \Fqstar  {{\Eff_q^*}}

\def \jsym#1#2{ {\left( {#1 \over #2} \right)} }
\def \csym#1#2{ {\left( {#1 \over #2} \right)} }
\def \qsym#1#2{ {\left( {#1 \over #2} \right)} }

\def \lg { \hbox{lg}\; }

\def \divides  { {\; \mid \;} }


\def  \AHU       {1}
\def  \BOS       {2}
\def  \COLL      {3}
\def  \CLSN      {4}
\def  \DF        {5}
\def  \EISB      {6}
\def  \EISA      {7}
\def  \EISN      {8}
\def  \EISE      {9}
\def  \FREI      {10}
\def  \GAU       {11}
\def  \HK        {12}
\def  \IR        {13}
\def  \JACV      {14}
\def  \JACU      {15}
\def  \KR        {16}
\def  \LEH       {17}
\def  \LEM       {18}
\def  \MULL      {19}
\def  \SW        {20}
\def  \SHA       {21}
\def  \SMIT      {22}
\def  \WEIL      {23}
\def  \WIKS      {24}
\def  \WILL      {25}
\def  \WH        {26}
\def  \WIN       {27}

\def  \normcomp  {1}
\def  \longdiv   {2}
\def  \csymflip  {3}
\def  \csymram   {4}
\def  \csymrho   {5}
\def  \csymminus {6}
\def  \regularity{7}
\def  \divns     {8}
\def  \charpoly  {9}
\def  \lengthlb  {10}
\def  \divbops   {11}
\def  \divbopslg {12}
\def  \logfact   {13}

\def  \normeqn   {15}
\def  \quadpar   {16}
\def  \qsymflip  {17}
\def  \qsymi     {18}
\def  \qsymram   {19}
\def  \qsymminus {20}

\centerline{On Euclidean Methods for Cubic and Quartic Jacobi Symbols}
\medskip
\centerline{Eric Bach and Bryce Sandlund}
\centerline{Computer Sciences Dept.}
\centerline{University of Wisconsin}
\centerline{Madison, WI 53706}

\bigskip
\centerline{July 18, 2018}

\bigskip
\noindent
{\bf Abstract.}
We study the bit complexity of two methods, related to the
Euclidean algorithm, for computing cubic and quartic analogs of
the Jacobi symbol.  The main bottleneck in such procedures is
computation of a quotient for long division.  We give examples
to show that with standard arithmetic, if quotients are computed naively 
(by using exact norms as denominators, then rounding), the algorithms
have $\Theta(n^3)$ bit complexity.  It is a ``folk theorem'' that
this can be reduced to $O(n^2)$ by modifying the division procedure.
We give a self-contained proof of this, and show that quadratic
time is best possible for these algorithms (with standard arithmetic
or not).  

We also address the relative efficiency of using reciprocity,
as compared to Euler's criterion, for testing if a given number is
a cubic or quartic residue modulo an odd prime.  Which is preferable
depends on the number of residue tests to be done.

Finally, we discuss the cubic and quartic analogs of Eisenstein's 
even-quotient algorithm for computing Jacobi symbols in $\Zee$.
Although the quartic algorithm was given by Smith in 1859,
the version for cubic symbols seems to be new.  As far as we know,
neither was analyzed before.   We show that both algorithms have 
exponential worst-case bit complexity.  The proof for the cubic algorithm
involves a cyclic repetition of four quotients, which may be of
independent interest.

\bigskip
\bigskip
\noindent
Presented as a poster at the 13th Algorithmic Number Theory
Symposium (ANTS-XIII), Madison, WI, July 16-20, 2018.

\bigskip
\bigskip
\noindent
Bryce Sandlund is now at the University of Waterloo.
E-mail addresses for authors: {\tt bach@cs.wisc.edu},
{\tt bcsandlund@uwaterloo.ca}

\vfill
\eject

\bigskip
\noindent
{\bf 1. Introduction.}

Let $\Eff$ be a finite field, of order $q$.  When
$e$ divides $q-1$, we can test whether $x$ is an $e$-th
power in $\Eff$ using Euler's criterion.  This is an ``exactness''
result, which asserts that, on $\Fqstar$, the image of the
$e$-th power map is the kernel of the $(q-1)/e$-th power map.
It therefore reduces the ``hard'' search
for a $y$ making $y^e = x$ to an ``easy'' evaluation
of $x^{(q-1)/e}$.

It is natural to ask whether exponentiation can be avoided.
It definitely can be, when $e=2$ and $q=p$, a prime
of $\Zee$.  This is because we can determine the quadratic
character by computing a Jacobi symbol, using a successive
division process similar to the Euclidean algorithm.
The correctness of this algorithm rests on the quadratic
reciprocity law, as extended by Jacobi to include composite
moduli.  For several such algorithms, see [\SHA].

In this paper we focus on $e=3,4$.
Since the results and methods are very similar in both cases, we will 
first restrict to $e=3$, and then take up $e=4$ toward 
the end of the paper.

When $e=3$, it is natural to move to the ring $\Zee[\rho]$, 
with $\rho = -1/2 + \sqrt{-3}/2 $ a primitive cube root of unity.  
When $\Fq$ is given to us as $\Zee[\rho]/(\pi)$
we can replace quadratic by cubic reciprocity, and produce
an analogous test that is free of exponentiation.  As far as we know,
the first complete description of such a test appeared in 1977,
in a paper by Williams and Holte [\WH].
They said nothing about its complexity, but later
Williams [\WILL] pointed out that the number of [arithmetic]
operations is logarithmic.  (See also M\"uller [\MULL].).
Later (2005), Damg{\aa}rd and Frandsen [\DF] pointed out that the bit
complexity is $O(n M(n))$, assuming we can multiply $n$-bit 
numbers with $M(n)$ bit operations.  If standard arithmetic is
used, this bound is $O(n^3)$.

Before describing our contributions we briefly mention 
some related work.  Damg{\aa}rd and Frandsen [\DF] gave Jacobi symbol
algorithms for $e=3,4$, related to the binary algorithm for the 
gcd in $\Zee$, that use $O(n^2)$ bit operations.  
Wikstr\"om [\WIKS] did the same for $e=8$.  Weilert found an algorithm
for the quartic Jacobi symbol with bit complexity $n^{1 + o(1)}$,
and suggested that a similar result could be proved for the cubic symbol 
[\WEIL, p.~145].  (We do not know if this was ever done.)
Jacobi symbol methods for prime $e \le 19$,
related to the Euclidean algorithm, were presented 
by Scheidler and Williams [\SW].  (Special attention was paid
to $e=5$.)

One goal of our paper is to give a complete self-contained discussion 
of the bit complexity of the Williams-Holte algorithm.  As far as we
know this has never been done.  In particular, we make the following 
observations.  First, the Damg{\aa}rd-Frandsen bound
is tight up to constant factors: we exhibit a sequence of 
input pairs that forces $\Omega(n M(n))$ bit operations.
Second, following ideas already used by Collins [\COLL] to 
compute the gcd in $\Zee[i]$, approximate quotients can be used in 
the Williams-Holte algorithm to obtain $O(n^2)$ bit complexity.
(Earlier, Kaltofen and Rolletschek [\KR] had computed gcd ideals
in rings of quadratic integers in quadratic time using a related 
but more complicated process.)
Finally, even if division could be done for free, the Williams-Holte
algorithm would still need $\Omega(n^2)$ bit operations in the worst case,
just to write down the remainders.

Our original motivation for this work was to investigate the 
preference, expressed by M\"uller and others, for the cubic Jacobi symbol 
algorithm over Euler's criterion.  In conjunction with this, we note
that there are really two different computational problems:
a) Given $p \in \Zee$, determine if $a$ is a cube mod $p$,
and b) Given $\pi \in \Zee[\rho]$, determine if $\alpha$ is
a cube mod $\pi$.  The situation is more complicated than appears
at first glance, and we will discuss it further in Section 6.

In Section 7, we discuss the cubic and quartic analogs
of an elegant but flawed algorithm of Eisenstein [\EISA] for computing 
the Jacobi symbol in $\Zee$.  The quartic version was given by 
Smith about 15 years after Eisenstein, but the cubic version may
well be new.  We show that both of these algorithms have exponential
bit complexity.  The proof for the cubic algorithm involves some
interesting dynamics: for certain input pairs, there is a repeating 
cycle of quotients, which has the effect of subtracting
constants from the pair.  Consequently, these intermediate
numbers decrease much more slowly than with the Williams-Holte
algorithm.

\bigskip
\noindent
{\bf 2. Mathematical Background.}

Let $\Zee$ denote the integers.  The ring of {\sl Eisenstein integers}
is $\Zee[\rho]$, where $\rho = -1/2 + \sqrt{-3}/2$ is a primitive cube 
root of unity.  It is the ring of algebraic integers of the imaginary
quadratic field $\Que(\rho)$.  Since $\rho^2 + \rho + 1 = 0$,
any Eisenstein integer can be written as $a + b \rho$, where
$a,b \in \Zee$.  The {\sl norm} of $\alpha = a + b \rho$
is
$$
N(\alpha) = \alpha \bar \alpha = a^2 - ab + b^2.
$$
When thinking of $\alpha$ as a complex number, we will write
$|\alpha| = \sqrt{N(\alpha)}$.

We can also think of the numbers $a + b \rho$, where $a$ and $b$
are real, as a 2-dimensional vector space over $\Ree$.  
Let $||\alpha||^2 = a^2 + b^2$.  From examining the quadratic
form $a^2 - ab + b^2$, we can derive the explicit 
norm-comparison inequality
$$
{2 \over 3} N(\alpha) \le  || \alpha ||^2 \le 2 N(\alpha).
\eqno{(\normcomp)}
$$

A strong version of long division holds in $\Zee[\rho]$: Given $\alpha$
and nonzero $\beta$, there are integral $q,\gamma$ for which
$$
\alpha = q \beta + \gamma,  \qquad N(\gamma) \le {3 \over 4} N(\beta).
\eqno{(\longdiv)}
$$
We can take $q = \lfloor \alpha / \beta \rceil$, where the brackets
indicate each coefficient is rounded to a nearest integer
[\IR, p.~13].  Consequently, $\Zee[\rho]$ is a 
principal ideal domain.
The units of this ring, the sixth roots of unity, are
$\pm 1, \pm \rho, \pm \rho^2$ (or, $\pm (1 + \rho)$).

A prime $p$ of $\Zee$ is ramified, split, or inert when
it is 0, 1, or 2 mod 3, respectively.  When $P$ is a
prime ideal of $\Zee[\rho]$, distinct from $(1-\rho)$,
the cubic symbol $\csym \alpha P$ is 
defined as the unique element of $\{0,1,\rho,\rho^2\}$ for which
$$
\csym \alpha P \equiv \alpha^{(NP - 1)/3} \pmod P .
$$
Here, $NP$ denotes the norm of $P$, that is, the size of 
$\Zee[\rho]/P$.
This extends to an analog of the Jacobi symbol, as follows.
Let $P = Q_1 \cdots Q_r$ be a factorization into such prime ideals, 
with repetitions allowed.  Then,
$$
\csym \alpha P = \prod_{i=1}^r \csym \alpha {Q_i} .
$$
For convenience, we also allow $r=0$ (that is, $P=(1)$),
in which case the symbol is 1.  (This follows [\BOS].)
This cubic symbol is asymmetric, as its arguments are a number
and an ideal.  To make it symmetric, we let $\csym \alpha \beta$ 
denote $\csym \alpha {(\beta)}$.

The cubic symbol obeys a reciprocity law for elements, but
it must be stated carefully, since multiplying the upper argument
by a unit can change the symbol's value.  Accordingly, we will
say that $\alpha$ is {\sl primary} if $\alpha \equiv \pm 1$ mod 3.
Note that a primary number cannot be divisible by $1-\rho$, and
the only primary units are $\pm 1$.  Some authors make a more
restrictive definition; to encompass this we let
{\sl 2-primary} mean $\equiv 2$ mod 3.  
Since $\Zee[\rho]/(3)^*$
is cyclic of order 6, each ideal prime to 3 has exactly one 2-primary
generator.

The {\sl cubic reciprocity} law consists of a main theorem and
two supplementary results.  First, suppose that
$\alpha,\beta$ are primary and relatively prime.
Then
$$
\csym \alpha \beta = \csym \beta \alpha
\eqno{(\csymflip)}
$$
Further, suppose $\beta = c + d\rho$, with $c = 3c' \pm 1$ and $d = 3d'$.
Then
$$
\csym {1-\rho} \beta = \rho^{\pm c'} = \rho^{-(c^2 - 1)/3}.
\eqno{(\csymram)}
$$
and
$$
\csym \rho \beta = \rho^{\mp(c'+d')} = \rho^{(c^2 - cd - 1)/3}    
\eqno{(\csymrho)}
$$
(The sign in (\csymram) matches the one in the definition of $c'$,
and the sign in (\csymrho) is its opposite.) For completeness we note that
$$
\csym {-1} \beta = 1,
\eqno{(\csymminus)}
$$
under the same condition.

As Collison [\CLSN] recounts, the history of cubic reciprocity makes 
for a complicated tale.  (See also [\FREI, \LEM, \SMIT].) 
Suffice it to say that (\csymflip)--(\csymrho) were first proved in print by 
Eisenstein in 1844 [\EISB, \EISN].  For a modern exposition of
these proofs, see Halter-Koch [\HK, \S 7.2].

We end this section with a few remarks on arithmetic in $\Zee[\rho]$.
$M(m,n)$ stands for the number of bit operations (bops) 
used to multiply an $m$-bit integer and an $n$-bit integer. 
We can assume that this depends only on input length.  Following
[\AHU, p.~280], we require $M(n) = M(n,n)$ to satisfy a 
regularity condition: if $a \ge 1$, then
$$
M(n) \le M(an) \le a^2 M(n).
\eqno{(\regularity)}
$$

The discrete binary logarithm (bit length function) is given by
$$
\lg n = \cases {
	1, & if $n=0$; \cr
        \lfloor \log_2 |n| \rfloor + 1, & if $n \ne 0$. \cr
        }
$$

By (\normcomp), if $\alpha = a + b \rho$, we have $a^2 + b^2 \le 2 N(\alpha)$.
So 
$$
\lg a, \lg b = O(\lg N(\alpha)).
$$
Therefore, we can compute $\alpha \pm \beta$ using
$O(\lg N(\alpha) + \lg N(\beta) )$ bops.  We can also
compute $\alpha \cdot \beta$ using 
$O(M(\lg N(\alpha), \lg N(\beta)))$ bops.  As for long division, 
we can take $q = \lfloor \alpha \bar \beta / N(\beta) \rceil$
and then $\gamma = \alpha - q \beta$; this will use
$M(\lg N(\alpha\beta))$ bops.  (The brackets indicate that
coordinates are rounded to nearest elements of $\Zee$.)
If we only want the
remainder, there is a formula of Jacobi [\JACV, pp.~249-250]: 
$$
\gamma = 
{ [\alpha \bar \beta \bmod N(\beta) ] \beta \over N(\beta) }
$$
(absolutely least remainders for the expression in brackets),
with the same complexity.  We will say more about long division
in Section 5.

\bigskip
\noindent
{\bf 3. Cubic Jacobi Symbols via Iterated Division.}

Just as in $\bf Z$, one can use continued fractions, or a variation
thereof, to compute cubic Jacobi symbols.  This was already known
to Jacobi.  In a paper from the autumn of 1837 [\JACU] that 
introduced the Jacobi symbol for $\Zee$, he noted that Gauss's 
method for computing the 
Legendre symbol $\jsym p q$ from the continued fraction of $p/q$ 
(presumably from [\GAU]) would still work with $p,q$ just relatively prime,
and then wrote (translation ours):
\smallskip
\item{}
     Exactly the same can be applied to biquadratic and cubic residues,
     for which I have introduced similar symbols.  The application of
     the symbols thus generalized affords, with some practice, agreeable
     relief.
\smallskip
\noindent
Although it is reassuring to read that even the great masters needed 
to hone their computational skills, we know of no place where Jacobi
explicitly wrote down such an algorithm for the cubic symbol.  
(In his number theory lectures [\JACV], delivered the preceding winter,
he sketched an algorithm for the quartic symbol employing
factorization, but said nothing directly about computing cubic symbols.)

Eisenstein mentioned a cubic symbol procedure in 1844 [\EISN, p.~32],
but gave no details.  He did present a full algorithm
for the quartic symbol a short time later [\EISE, \S 10].  In
the cubic case, full disclosure apparently had to wait until 1977, 
when Williams and Holte [\WH] 
wrote up the details for an algorithm that we now recognize as
the cubic analog of Eisenstein's quartic method.
We now give their algorithm, with an extension that allows for 
inputs that are not relatively prime.

\smallskip
\item{1.} [Inputs.]  
Let $\alpha = a + b \rho$ and $\beta = c + d\rho$, with 
$c \not\equiv 0$, $d \equiv 0$ mod 3.  (That is, $\beta$ is primary.)

\smallskip
\item{2.} [Base cases.]  
If $\beta = \pm 1$ or $\alpha = \pm 1$, return 1 and stop.

\smallskip
\item{3.} [Find remainder.]  
Let $q = \lfloor \alpha / \beta \rceil$.  Compute
$$
\gamma = e + f \rho = \alpha - q \beta .
$$
If $\gamma=0$, return 0 and stop.

\smallskip
\item{4.} [Remove ramified factors.]  
Find the least $m \ge 0$ for which
$$
\gamma' = e' + f' \rho = {e + f \rho \over (1 - \rho)^m}
$$
is not divisible by $1-\rho$. 
(That is, has $e' + f' \not\equiv 0$ mod 3.)

\smallskip
\item{5.} [Make primary.]
Find $n$ with $0 \le n < 3$ so that
$$
\gamma'' = {e' + f' \rho \over \rho^n}
$$
is primary.

\smallskip
\item{6.} [Recursion.]
Return the right side of
$$
\csym \alpha \beta 
=
\rho^{-m(c^2-1)/3 + n (c^2 - cd - 1)/3}
\csym \beta {\gamma''}
$$

The algorithm must terminate because $N(\beta)$ strictly decreases.
We briefly indicate why the output is correct.  When $\beta$ is a unit,
the value of the symbol is 1, and if not, but with $\alpha = \pm 1$,
the same is true.  So let there be at least one division step.  Using
indexing to denote recursion depth, we have
$$
\def \sp {\qquad\quad}
\eqalign{
        \alpha_1 & = q_1 \beta_1 + \gamma_1
        \sp        = q_1 \beta_1 + (1-\rho)^{m_1} \rho^{n_1} \gamma''_1, \cr
        \alpha_2 & = q_2 \beta_2 + \gamma_2
        \sp        = q_2 \beta_2 + (1-\rho)^{m_2} \rho^{n_2} \gamma''_2, \cr
                 & \vdots \cr
        \alpha_{t-1} & = q_{t-1} \beta_{t-1} + \gamma_{t-1}
                       = q_{t-1} \beta_{t-1} 
                       + (1-\rho)^{m_{t-1}} \rho^{n_{t-1}} \gamma''_{t-1}, \cr
        \alpha_t      & = q_t \beta_t, \cr
        }
\eqno{(\divns)}
$$
where $\alpha_2 = \beta_1, \beta_2 = \gamma_1''$, and so on.  (We imagine
the last division is done, whether needed or not.)
Since $\beta$ is primary, $\gcd(\alpha,\beta)$
is coprime to $1-\rho$, so (using induction), $\beta_t$ is a gcd of
$\alpha, \beta$.  If $N(\beta_t)>1$, the last call returns 0, which
becomes, correctly, the output.  If $N(\beta_t) = 1$ (that is, $\gamma''_{t-1}$ 
is $\pm 1$), the $t$-th call returns 1, and correctness of the output
follows from (\csymflip)--(\csymrho).

Note that the algorithm computes a gcd, free of charge, in cases where 
the symbol is 0.  This is similar to the situation in $\Zee$
[\SHA, p.~608].

Let us now discuss some implementation details.  Williams and Holte
computed the ratio in Step 3 as
$$
{\alpha \over \beta}
=
{\alpha \bar \beta \over N(\beta)}.
$$
In Step 4, we can divide by $1-\rho$ using the formula
$$
{e + f \rho \over 1 - \rho} = {2e - f \over 3} + {e+f \over 3} \rho;
$$
the result is integral iff 3 divides $e+f$.  Regarding Step 5,
division by $\rho$ (that is, multiplication by $\rho^2$)
permutes
$$
e + f \rho, \quad (f-e) - e \rho, \quad -f + (e-f)\rho
$$
in cyclic order.  Examination of cases shows that exactly one of
these is primary.  Finally, the numerators in the exponent in Step 6 
need only be computed mod 9.  Williams and Holte gave this exponent
in the equivalent form $(2m+n)(c^2-1)/3 - n cd/3$.

\bigskip
\noindent
{\bf 4. A ``Bad'' Sequence of Input Pairs.}

Our goal is to construct a sequence of inputs for which the Williams-Holte
algorithm needs $nM(n)$ bit operations.

Consider the recurrence relation
$$
\eqalign{
    \xi_0  &= -1, \cr
    \xi_1  &=  2, \cr
    \xi_n  & = 3 \rho \xi_{n-1} + \xi_{n-2}, \qquad n \ge 2. \cr
    }
$$
This defines a sequence of 2-primary integers in $\Zee[\rho]$, because
all $\xi_n$ are 2 mod 3, as can be proved by
induction.  The first few sequence elements are
$$
\def \hf {\hfill}
\matrix {
           n   & \  \xi_n                    & N(\xi_n)    \cr
               &                           &           \cr
	   0   & \hf -1                    & \hf 1 \cr 
	   1   & \hf 2                     & \hf 4 \cr 
	   2   & \hf -1 + 6 \rho           & \hf 43 \cr 
	   3   & \hf -16 -21 \rho          & \hf 361 \cr 
	   4   & \hf 62 + 21 \rho          & \hf 2983 \cr 
	   5   & \hf -79 + 102 \rho        & \hf 24703 \cr 
	   6   & \hf -244 - 522 \rho        & \hf 204652 \cr 
	   7   & \hf 1487 + 936 \rho       & \hf 1695433 \cr 
	   8   & \hf -3052 + 1131 \rho     & \hf 14045677 \cr 
	   9   & \hf -1906 - 11613 \rho     & \hf 116360227 \cr 
           10  & \hf 31787 + 30252 \rho    & \hf 963976549 \cr 
         }
$$

We now show that these grow rapidly.  The
characteristic polynomial for the recurrence relation 
is
$$
    X^2 -  3 \rho X - 1,
\eqno{(\charpoly)}
$$
with zeroes
$$
\lambda,\mu = {3 \rho \pm \sqrt{9 \rho^2 + 4} \over 2}.
$$
(Incidentally, $\lambda \not\in \Que(\rho)$ since (\charpoly)
is irreducible modulo any prime divisor of 19.)  Numerically, 
$$
\matrix{
    \lambda \doteq -1.7059 + 2.3182 i,         & |\lambda| \doteq 2.8783, \cr
    \mu\   \doteq \phantom{-}0.2059 + 0.2798 i,& |\mu| \doteq  0.3474. \cr
    }
$$
By the theory of linear recurrence relations, there are
numbers $A$ and $B$ for which
$$
    \xi_n = A \lambda^n + B \mu^n.
$$
(Numerically, $A \doteq -0.4669 - 0.6442 i, B \doteq -0.5331 + 0.6442 i$.)

For large $n$ the second term is insignificant, and we have
$$
    \log( N( \xi_n ) ) \sim n \log ( N(\lambda) ) \doteq 2.1144 n .
$$
So, in norm, the length of $\xi_n$ grows linearly, and the number
of recursive calls for the algorithm on the inputs $\xi_n, \xi_{n-1}$
is $n$.

We now show that applying the algorithm with $\alpha = \xi_n$ and
$\beta = \xi_{n-1}$ is tantamount to running the recurrence backwards. 
Observe first that if $a + b \rho = x + yi$, we have
$$
\eqalign{
         a & =                x +    {1 \over \sqrt 3} y, \cr
         b & = \hbox{\phantom{x +}}  {2 \over \sqrt 3} y. \cr
         }
$$
So for large $n$,
$$
\xi_n / \xi_{n-1} \sim \lambda \doteq -0.3675 + 2.6769 \rho,
$$
which rounds to $3 \rho$.  (Apparently $n\ge 3$ is large enough.)
Therefore, the algorithm will produce
$$
\gamma = \xi_n - 3\rho \xi_{n-1} = \xi_{n-2}
$$
for the next stage.

If $\xi_n = a_n + b_n \rho$, it can be shown that the vector
$$
v_n = \pmatrix { a_n \cr 
                 b_n \cr
               }
$$
satisfies the recurrence $v_n = M v_{n-1} + v_{n-2}$, where 
$M$ is a $2 \times 2$ matrix.  The question of the lengths of 
$a_n$ and $b_n$ is more delicate, as they don't always increase.  
This is already apparent from the indicated values.

To get around this problem, we note that when $N(a + b \rho)$ is large, 
at least one of $a$ or $b$ is large.  Indeed, by (\normcomp),
$ a^2 + b^2 \ge {2 \over 3} N(a + b \rho)$, so
$$
\max\{ |a|, |b| \} \ge \left( N(a+b\rho)/3 \right)^{1/2}.
\eqno{(\lengthlb)}
$$

Suppose we implement the Williams-Holte algorithm by first computing 
an exact fractional quotient $\alpha \bar \beta / N(\beta)$,
using the usual norm formula.  If the inputs are $\xi_n,\xi_{n-1}$,
just computing the denominators alone will use $\Theta(n)$
multiplications in $\Zee[\rho]$, of numbers whose norms decrease 
geometrically over time.  At the halfway point, we must still square
a integer whose length is $\Omega(n)$, by (\lengthlb).
Since (\regularity) implies $M(n/c) \ge M(n)/c^2$, we have
already used $\Omega(nM(n))$ bops by this point.  

In particular, with standard arithmetic, the worst-case bit complexity
of the algorithm is cubic: on the inputs $\alpha,\beta$ it is 
$\Theta(\lg N(\alpha \beta))^3$.

Instead of the standard norm formula, we could use others, such as
$(a-b)^2 + ab$ or $((2a-b)^2 + 3 b^2)/4$.  
However, any reasonable alternative
would lead us to the same conclusion.  Let us restrict attention to
algorithms that compute the norm as
$$
{1 \over s} \sum_{i=1}^k (t_i a + u_i b)(v_i a + w_i b)
$$
where $s, t_1, \ldots, w_k$ are fixed integers.  (Algebraic complexity
theory supports this choice [\WIN, \S IIIc].) 
If $\ell = \max\{\lg |a|, \lg |b|\}$, then 
$$
\lg(t_i a + u_i b), \lg(t_i a + u_i b) \le \ell + O(1).
$$
On the other hand, $\lg (N(a+ b\rho)) \ge 2 \ell - O(1)$, by
(\normcomp).   We must multiply at least
one pair of integers whose lengths are $\ge \ell/2$, for if not,
the sum on the right hand side would have length $\le 3\ell/2 + O(1)$,
and this would still hold after division by $s$.
(Recall that $k$ is fixed.)  Since $M(\ell/2) \ge M(\ell)/4$,
this is $\Omega(M(\ell))$ bops.

Incidentally, the minimum value of $k$ is 2.  To see this, observe
that any one-multiplication algorithm that yields $a^2 - ab + b^2$ could
be normalized into $(a + ub)(a + wb)$, with $u,w \in \Que$.  This
forces $uw=1, u+w = -1$, which is a system with no real solutions. 

\bigskip
\noindent
{\bf 5. A Quadratic Bit Complexity Bound.}

This section provides a complete analysis for the
the bit complexity of an approximate-division version of
the Williams-Holte algorithm.  Many of the techniques
we use reflect previous analyses of Euclidean gcd algorithms, so
the quadratic bit complexity bound is probably a ``folk
theorem.''  Nevertheless, we thought it worthwhile to record
a proof, because no reference known to us does this.

Let us first review the standard Newton iteration for
computing $\beta^{-1}$, where $\beta$ is a nonzero
complex number.  This iteration is
$$
\xi' = \xi(2 - \beta \xi).
$$
Suppose that $\xi = (1 - \epsilon)/\beta$ with $|\epsilon|<1$.
Then
$$
\xi' = {1 - \epsilon \over \beta} \left( 1 + \epsilon \right)
   = {1 - \epsilon^2 \over \beta}.
$$
By induction, the $i$-th iterate will be
$$
\xi_i = {1 - \epsilon^{2^i} \over \beta}.
$$
(The starting value is denoted $\xi_0$.)  To obtain relative error
$\le 2^{-m}$, we need $|\epsilon|^{2^{i}} \le 2^{-m}$, which we
can solve to obtain
$$
i \ge \log m - \log\log |\epsilon|^{-1}.
$$
(Here, the logs are binary.)
With the usual expedient of doubling the precision at each step,
this will require $O(M(m))$ bops.  Here we are assuming that
$|\epsilon|$ belongs to a compact subset of the unit interval
$(0,1)$; the implied constant will depend on this subset.

We now consider a particular starting point.  Let
$\beta = c + d \rho$ be a nonzero Eisenstein integer.
Then, there are $r,s$ making $2^{r-1} \le |c| < 2^r$ and 
$2^{s-1} \le |d| < 2^s$.  (If $c=0$ we take $r=-\infty$,
and similarly for $d$.) In $\Que(\rho)$, the exact inverse is
$$
\beta^{-1} = {c - d \over N(\beta)} - {d \over N(\beta)}\rho.
$$
Then, with $e = 2 \max\{r,s\} + 2$,
the starting point
$$
\xi_0 = {c - d \over 2^e} - {d \over 2^e}\rho
$$
satisfies $\xi_0 = \eta \beta^{-1}$ with $1/32 \le \eta \le 3/4$.  
(To prove this, observe that 
$(c^2+d^2)/2 \le 2^{2 \max\{r,s\}} \le 4(c^2 + d^2)$,
and then use (\normcomp) to get $(4/3) N(\beta) \le 2^e \le 32 N(\beta)$.)
Since $\epsilon = 1-\eta$, this implies $1/4 \le \epsilon \le 31/32$.

We now consider the bit complexity of long division.
Let us argue informally.  The coefficients of any number in $\Que(\rho)$
can be written in base 2, and if these parts are aligned,
each bit pair can be chosen in four possible ways.  Thus,
$35/2+3\rho$, equal to $1001.1 + 11 \rho$ in base 2, becomes
$
\def \pmtrx#1#2{ ({#1 \atop #2}) } 
\pmtrx 1 0 \pmtrx 0 0 \pmtrx 0 1  \pmtrx 1 1
.
\pmtrx 1 0  \pmtrx  0 0  \pmtrx 0 0  \ldots
$,
or, using a straightforward encoding, $2013.200\cdots$\ .

Let $\alpha,\beta \in \Zee[\rho]$ be the inputs for long
division.  Let these, along with the exact inverse of $\beta$,
have the representations:
$$
\eqalign{
\alpha &= \overbrace{******\cdots***}^{\hbox{$k$ digits}}.000\cdots \cr
\beta  &= \phantom{***}\overbrace{***\cdots***}
             ^{\hbox{$\ell$ digits}}.000\cdots \cr
\beta^{-1} &= \phantom{******\cdots***}.\overbrace{000\cdots000}^
                               {\hbox{$\ell$ digits}}
                               \overbrace{**\cdots**}^
                               {\hbox{$m$ digits}}
                               {*** \cdots} \cr
}
.
$$
Since the quotient is 0 if $k<\ell$, we can assume $\ell \le k$.  
To obtain the rounded quotient, numerical inversion of 
$\beta$ must deliver $m = k-\ell + O(1)$ digits past the leading
$\ell$ 0's.  For Newton's method with precision
doubling, as indicated above, this uses $O(M(m))$ bops.
Multiplying this approximate inverse by $\alpha$ 
and then rounding the result to $q$ uses $O(M(m,k))$ bops.
(One can show that if the error in each computed coefficient of
$\alpha \beta^{-1}$ is $\le \delta$, the constant in (\longdiv) is
increased only slightly, to $3/4 + 3 \delta + O(\delta^2)$.) 
Finally, we obtain $q \beta$ with $O(M(m,\ell))$ bops.  This totals to
$$
O( M(k,m) +  M(m) + M(\ell,m) )
\eqno{(\divbops)}
$$
There is one further subtraction to produce $\gamma$, which
we will ignore since it will use no more work than any bound
we prove.

For standard arithmetic, $M(m,n) = O(mn)$ so (\divbops) becomes
$$
O( km + (m+\ell)m ) = O(km) = O(\lg N(\alpha) \lg N(q) ).
\eqno{(\divbopslg)}
$$
As a worst-case bound, this is not impressive, since the quotient
could be almost as long as the dividend.  However, for a Jacobi
symbol calculation, not all quotients can be large.  This makes
it useful to have the quotient length as a factor.

Our final goal in this section is a quadratic bit complexity bound
for the Williams-Holte algorithm, with long division implemented
in the above manner.  This will be achieved using only standard arithmetic.
As we will see this bound cannot be improved.  

Since the norms can be compared within our time bound, we will assume that 
$N(\alpha) \ge N(\beta)$.  By (\longdiv) the maximum recursion depth $t$
(from (\divns)) is $O(\lg N(\beta))$.
Now we consider each part of the algorithm separately.  

First, we consider the work for long division (Step 3).  We first note that
$$
\log(x+1) \le \log(x) + 1
\eqno{(\logfact)}
$$
(binary logs), provided that $x \ge 1$.  This can be proved by 
comparing derivatives.  Using (\divbopslg), the number of bops
in the long divisions is at most a constant times
$$
\eqalign{
\sum_{i=1}^t \lg N(\alpha_i) \lg N(q_i)
       &\le \lg N(\alpha) \sum_{i=1}^t (2 \log |q_i| + 1) \cr
       &\le \lg N(\alpha) \sum_{i=1}^t (2 \log |\alpha_i/\beta_i| + 3) \cr
       &\le \lg N(\alpha) (2 \log |\alpha_1/\beta_t| + 3t) \cr
       & = O( \lg N(\alpha) )^2, \cr
       }
$$
using the triangle inequality and (\logfact).

Next we consider Step 4.  Let us call a 
division by $1-\rho$ {\sl successful} if the quotient is
integral, and {\sl unsuccessful} if it is not.  There are $\le t$
unsuccessful divisions, and since $N(1-\rho)>1$, 
$O(\lg N(\beta))$ successful ones.  The 
work for each such division is linear, so Step 4 accounts for 
$O(\lg N(\beta))^2$ bops.

The same is true of Step 5, since each invocation of it uses
$O(\lg N(\beta))$ bops.

Finally, for Step 6, we assume $c$ and $d$ are reduced mod 9 before 
computing the exponent.  This uses $O(\lg N(\beta))$ bops.
The multiplier $n$ is bounded, and it is simplest conceptually
to imagine multiplication by $m$ as repeated addition, since the 
sum of all such multipliers is $O(\lg N(\beta))$.  Then,
Step 6 will account for $O( \lg N(\beta))^2$ bops.

Adding the last four bounds and using $N(\beta) \le N(\alpha)$
gives us $O(\lg N(\alpha))^2$, as claimed.

To end this section, we point out that the quadratic bit complexity 
estimate is best possible for this algorithm.  Indeed, it holds
for any varition of it that computes sufficiently accurate quotients.  
To see this, suppose the inputs are $\alpha=\xi_n$ and $\beta=\xi_{n-1}$, 
as defined in Section 4.  
The ratio of two successive $\xi_i$'s is about $-0.37 + 2.7\rho$.
If their quotient is computed accurately enough, there will
be no ambiguity about the rounded coefficients, since
the rounding can only have more than one possible result
if a coefficient is an integer multiple of 1/2.
Therefore, the approximate-quotient algorithm will ``track'' the
exact algorithm for all but the last few steps (and produce
primary remainders).  Just to write down the successive
remainders will require $\Omega(\lg N(\alpha))^2$ bops.

Assuming Step 4 searches by incrementing $m$, we can
can also force a linear number of divisions here,
by choosing $\alpha = 3^m + (1-\rho)^m + 1$ and $\beta = 3^m+1$, 
which is primary.  Then, $\alpha/\beta = 1 + (1-\rho)^m / (3^m + 1)$
which rounds to 1 since $|(1-\rho)^m / (3^m + 1)| < 3^{-m/2}$.
At the top level, the number of divisions by $1-\rho$ in Step 4
will be $m+1$, with the last (unsuccessful) one leaving $\gamma'' = 1$.
With these inputs (at least with binary representation),
Step 4 alone would consume $\Omega( \lg N(\alpha))^2$ bops.

\bigskip
\noindent
{\bf 6. Cubic Residue Tests.}

The reciprocity law suggests a procedure for determining if 
$a \in \Zee$ is a cubic residue of the prime $p \in \Zee$ [\MULL].  
We need only consider $p \equiv 1$ mod 3, and assume $a$ is prime to $p$.
The first step is to solve the norm equation
$$
p = x^2 - xy + y^2
\eqno{(\normeqn)}
$$
for integers $x,y$.  It is sufficient to solve the related equation
$$
p = s^2 + 3 t^2
\eqno{(\quadpar)}
$$
and then set $x = s+t, y = 2t$.
To solve (\quadpar), we can first solve $s^2 + 3 \equiv 0$ mod $p$,
and then determine $t$ from the ordinary continued fraction
expansion for $s/p$.  See [\WILL, p.~366] for details of this.
Using standard algorithms for the square root mod $p$,
the bit complexity of this is $O(nM(n))$; the expensive
part is finding $s$.  By (\normeqn), one of the prime ideals 
$P$ lying above $p$ is generated by $\pi = x+\rho y$, which
we can make primary if it is not already.  The cubic Jacobi
symbol $\csym a \pi$ is 1 iff $a$ is a cubic residue mod $p$.

Whether this beats Euler's criterion depends on several factors.
If we test many $a$ against one $p$, then the effort of computing
$\pi$ is, in effect, amortized over the sequence of $a$'s.  So
the procedure is definitely worth using, once we have about
$\log p$ values of $a$.  For a ``one-off'' cubic residue test,
Euler's criterion should be used, since
all the most efficient algorithms to find square roots mod $p$ use 
exponentiation anyway.  Before digital computers, however, 
quadratic partitions such (\quadpar) were systematically tabulated.  
(Jacobi seems to have been the first to do this; see the 1846 
version of [\JACU].) With this representation of $p$ in hand, 
the advantage then shifts to the cubic Jacobi algorithm, assuming 
long division is implemented efficiently.

As an aside, we note that for $p \equiv 1$ mod 3, finding 
$\sqrt{-3}$ mod $p$ reduces
to solving the norm equation for $p$.  For, suppose we know
integers $x,y$ making $x^2 - xy + y^2 = p$.  Then we have
$$
\left({x+y \over 2}\right)^2 + 3 \left({x-y \over 2}\right)^2 = p,
$$
so that
$$
z = {x+y \over x-y}
$$
is a square root of $-3$ mod $p$.  The denominator cannot
vanish mod $p$; if it did, then from $(x-y)x + y^2 = p$
we could conclude $p \mid y^2$ and then $p | y$, which contradicts
$x^2 + y^2 \le 2 N(x + \rho y) = 2p$ (a consequence of (\normcomp)).

We end this section with a few remarks on the tabulation 
and computation of solutions to (\quadpar).  By 1925, Cunningham had 
tabulated the (essentially unique) solutions for all eligible primes up
to $p = 125683$.  Lehmer's 1941 survey of number-theoretic tables
[\LEH] mentioned nothing better, so this record likely still stood
at the dawn of the computer age.   It is not clear to
us if anyone went further, but it is worth noting that a list
for all such $p$ up to $N$ could be made, in amortized
polynomial time, by looping through all positive $x \le \sqrt N$ and 
$y \le \sqrt{N/3}$, combining their squares, sorting to 
remove duplicates, and then comparing to a list of primes, obtained
from the sieve of Eratosthenes.

\bigskip
\noindent
{\bf 7. The Quartic Jacobi Symbol.}

The results we have proved generalize to testing for fourth powers
in finite fields.  Since the arguments are so similar, we will
be brief.  The ring of {\sl Gaussian integers} is $\Zee[i]$; 
this is a principal
ideal domain with units $\pm 1, \pm i$.  Here, the norm coincides 
with the square of Euclidean length: $N(a+bi) = a^2 + b^2$.
Again, we have division with remainder: (\longdiv) still holds,
but with $3/4$ replaced by $1/2$ [\IR, p. 12].
Only $p=2$ is ramified; primes $p \equiv 1$ mod 4 are split and
primes $p \equiv 1$ mod 4 are inert.

When $P \ne (1+i)$ is a prime ideal, the quartic character is
defined by the condition
$$
\qsym \alpha P \equiv \alpha^{(NP-1)/4} \bmod P;
$$
we define a Jacobi symbol for $\Zee[i]$ just as before.

Finally, {\sl primary} means $\equiv 1$ mod $(1+i)^3$.  Then,
$a+bi$ is primary iff $2 \divides b$ and $a+b \equiv 1$ mod 4
[\IR, p.~121].

The replacements for (\csymflip)--(\csymminus) are as follows.
Let $\alpha = a+bi$ and $\beta= c+di$ be primary.  Then
$$
\csym \alpha \beta = (-1)^{(a-1)(c-1)/4} \csym \beta \alpha,
\eqno{(\qsymflip)}
$$
$$
\csym {1+i} \beta = i^{(c-d-d^2-1)/4},
\eqno{(\qsymram)}
$$
$$
\csym i \beta = i^{-(c-1)/2},
\eqno{(\qsymi)}
$$
and
$$
\csym {-1} \beta = (-1)^{(c-1)/2}.
\eqno{(\qsymminus)}
$$
See Halter-Koch [\HK, \S 7.3-7.4] for proofs.

A quartic Jacobi symbol algorithm, entirely analogous to the cubic one
in Section 3 (including the possibility that $\gcd(\alpha,\beta) \ne 1$)
was given by Eisenstein in 1844 [\EISE, \S 10].  For compatibility with
Section 3, we express the division step as
$$
\alpha = q \beta + \gamma
       = q \beta + (1+i)^m i^n \gamma'',
$$
where $N(\gamma) \le N(\beta)/2$ and
$\gamma'' = e'' + f''i$ is primary.  The algorithm of Section 3
can then be modified as follows.  Step 1 should return 1 on the
condition that $\beta=1$ or $\alpha=1$.
Steps 4 and 5 are the same, but with $1-\rho$ and $\rho$ replaced
by $1+i$ and $i$, where $0 \le n < 4$. Finally, the recursion becomes
$$
\qsym \alpha \beta = 
i^{m(c-d-d^2-1)/4 - n(c-1)/2}
(-1)^{(e''-1)(c-1)/4}
\qsym \beta {\gamma''}.
$$

As before, suppose that we compute $q$ by computing
$\alpha \bar \beta / N(\beta)$ exactly, and then rounding coefficients.
Consider the sequence defined by the recurrence relation 
$\xi_n = 2(i+1)\xi_{n-1} + \xi_{n-2}$, with initial conditions 
$\xi_0 = 1, \xi_1 = 5$.  The dominant root of its characteristic polynomial 
$X^2 - 2(i+1)X - 1$ is $(i+1) + \sqrt{2i+1} \doteq 2.2720 + 1.7862i$.
This makes the sequence grow rapidly, and for large $n$,
the coordinatewise nearest Gaussian integer to $\xi_n/\xi_{n-1}$ is $2+2i$.
Therefore, arguing as we did in Section 4, Eisenstein's algorithm
will use $\Omega(n M(n))$ bops in the worst case.

As we did for the cubic algorithm, we can obtain a quadratic bit-complexity
bound with standard arithmetic by using Newton iteration to determine
quotients for long division.  Our starting point for inverting 
$\beta = c+di$ is 
$$
\xi_0 = {c \over 2^e } 
      - {d \over 2^e } i,
$$
where $e$ is defined from $c$ and $d$ just as in 
Section 5.  As before, $2 N(\beta) \le 2^e \le 16 N(\beta)$, which
gives $1/2 \le \epsilon \le 15/16$.  Again, a quadratic bound on 
bops is best possible for Eisenstein's quartic algorithm.

Finally, all of the conclusions in Section 6 hold true for
testing quartic residues modulo primes $p \equiv 1$ mod 4.  
In particular, we can reduce computing
$\sqrt {-1}$ mod $p$ to solving the Diophantine equation $p = x^2 + y^2$
by observing that in such a representation, $p$ divides neither $x$
nor $y$. So $(y/x)^2 \equiv -1$

\bigskip
\noindent
{\bf 8. Generalizing the Even-Quotient Algorithm.}

The method most commonly presented in textbooks for computing the
Jacobi symbol in $\bf Z$ could be called an even-to-odd remainder 
algorithm, because a maximal power of 2 is removed from the remainder 
before the recursive call.  There is another algorithm, published 
by Eisenstein in [\EISA], in which the
{\sl quotients} are even.  This algorithm uses division steps of
the form
$$
\alpha = q \beta + \gamma,
$$
in which $q$ is even, $|\gamma| < \beta$, and $\gamma$ is odd.
Eisenstein called it a ``simpler'' algorithm, but it is actually
more complex, as far as worst-case behavior is concerned.  In
particular, when applied to consecutive odd numbers, the number
of division steps is exponential in bit length [\SHA, p.~608].

The even-quotient idea generalizes to both cubic and quartic symbols.
The cubic version seems not to have been discussed before, so we
treat that first.  Let $N(\beta) > 1$.  Then, if $\alpha$ is any
other element of $\Zee[\rho]$, we can find a $q$, divisible
by $1-\rho$, for which
$$
\alpha = q \beta + \gamma,   \qquad N(\gamma) < N(\beta).
$$

To prove this, consider a nonzero $\zeta \in \Que(\rho)$.  We want to write
$\zeta = q + \omega$ with $q \equiv 0$ mod $1-\rho$ and $|\omega|<1$.  
Form $q$ in the usual way, by rounding coefficients of $\zeta$ 
to nearest integers.  This makes $|\omega|<1$.  If $q$ is not 
divisible by $1-\rho$, consider choosing a sixth root of unity 
$\epsilon$ and replacing $q$ by $q + \epsilon$.  As we go around 
the unit circle, the sixth roots of unity, reduced mod $1-\rho$, 
alternate between 1 and 2.  Therefore, three of the choices for 
$\epsilon$ will make $q$ divisible by $1-\rho$.  Let the new 
remainder be $\omega' = \omega - \epsilon$.  We now show that 
$|\omega'|<1$ is still possible.  If the angle $\theta = \arg \omega$
lies between 0 and $\pi/3$, we have $|\omega-1| < 1$.  (To see this, 
observe that there are two congruent equilateral triangles, with 
vertices $0, 1, -\rho^2$ and $-1, 0, \rho$.)
By symmetry, we also have $|\omega - (-\rho^2)|<1$.  The same argument
can be made for each of the other five sectors.  The desired result 
then follows upon taking $\zeta = \alpha/\beta$.

A cubic analog to Eisenstein's even-quotient algorithm can then be
defined so as to use (\divns), but with all $q_i \equiv 0$ mod $1-\rho$,
and all $m_i=0$.  (That is, the division in Step 3 has an additional
constraint, and Step 4 is deleted from the algorithm.)

Just as for the algorithm in $\Zee$, the worst-case bit complexity is
exponential.  As far as we know, this is not easy to discover by
manipulating formulas.  (Some fruitless effort went into this quest.)
Instead, we turned to the ``tools of ignorance.''  After implementing
the algorithm in Maple, we ran it on all possible input coefficients
in $[0,10]$.  Keeping track of ``record values'' for the number of
iterations revealed that for $\alpha = (3k+2)\rho$, and
$\beta = 1 + (3k+3)\rho$, the number of iterations is $4k+3$.
For these inputs, the bit length $n$ is $\sim 2 \log_2 k$, making
the number of iterations about $2^{n/2}$.

To prove that our observation would always hold, we noted that after
two ``warmup'' iterations, the algorithm settled into a repeating
cycle of four quotients:
$$
-2 - \rho, \ -1 - 2\rho, \ 2+\rho, \ -2 - \rho.
$$
We then executed the algorithm symbolically, using Laurent series
around $k=0$ (that is, expansion in powers of $1/k$) to guarantee
that rounding would behave correctly for all sufficiently large $k$.
Doing this verified that the observed number of iterations holds
holds for all but a finite number of $k$, from which the exponential 
bit complexity follows.

In 1859, Smith [\SMIT, p.~87] presented the analogous algorithm
for computing quartic Jacobi symbols.  He asserted without proof
that the corresponding long division (i.e., with all quotients
divisible by $1-i$) is possible.  That can be proved with a similar
argument; it is only necessary to observe that all powers
of $i$ are 1 mod $i+1$, and then consider quarter-circular sectors
with angles centered on multiples of $\pi/2$.

It seems to have escaped notice that the worst-case bit complexity
of this algorithm is also exponential.  If we start with inputs
$4m+1, 4m-3$, there could be about $m$ division steps, since
$$
4m + 1 = 2(4m-3) - (4m-7).
$$
(Incidentally, this is also a bad example for Eisenstein's algorithm.)
To be fair to Smith, he did point out that Eisenstein's version of the 
quartic algorithm ``terminates more rapidly'' than his did, but 
without making this quantitative.

\bigskip
\noindent
{\bf Acknowledgments.}
The authors were supported by the National Science Foundation
(CCF-1420750).

\bigskip
\noindent
{\bf References.}

\item{\AHU.}
A. V. Aho, J. E. Hopcroft, and J. D. Ullman,
The Design and Analysis of Computer Algorithms,
Addison-Wesley, 1974.

\item{\BOS.}
W. Bosma,
Cubic reciprocity and explicit primality tests for $h \cdot 3^k \pm 1$,
in A. van der Poorten, A. Stein, eds., 
High Primes and Misdemeanours: Lectures in Honour of the 60th Birthday 
of Hugh Cowie Williams (Fields Institute Communications v. 41), AMS,
2004.

\item{\COLL.}
G. E. Collins,
A fast Euclidean algorithm for Gaussian integers,
J. Symbolic Comput. 33, 385-392, 2002.

\item{\CLSN.}
M. J. Collison, The origins of the cubic and biquadratic
reciprocity laws, Arch. Hist. Exact Sci., v. 17, pp.~63-69, 1977.

\item{\DF.}
I. B. Damg{\aa}rd 
and G. S. Frandsen,
Efficient algorithms for the gcd and cubic residuosity in
the ring of Eisenstein integers, J. Symbol. Comput. 39, 643-652, 2005.

\item{\EISB.}
G. Eisenstein, 
Beweis des Reciprocit\"atssatzes f\"ur die cubischen Reste
in der Theorie der aus dritten Wurzeln der Einheit
zusammengesetzen complexen Zahlen, 
J. reine angew. Math. 27, 289-310, 1844.

\item{\EISA.}
G. Eisenstein, 
Einfacher Algorithmus zur Bestimmung des Werthes von $\jsym a b$,
J. reine angew. Math. 27, 317-318, 1844.

\item{\EISN.}
G. Eisenstein, 
Nachtrag zum cubischen Reciprocit\"atssatze fur die aus dritten
Wurzeln der Einheit zusammengesetzen complexen Zahlen.
Criterien des cubischen Characters der Zahl 3 und ihrer Theiler,
J. reine angew. Math. 28, 28-35, 1844.

\item{\EISE.}
G. Eisenstein, 
Einfacher Beweis un Verallgemeinerung des Fundamentaltheorems
f\"ur die biquadratischen Reste,
J. reine angew. Math. 28, 223-245, 1844.

\item{\FREI.}
G. Frei, The reciprocity law from Euler to Eisenstein,
in S. Chikara, S. Mitsuo, and J. W. Dauben, eds., The
Intersection of History and Mathematics, Birkh\"auser, 1994.

\item{\GAU.}
C.F. Gauss,
Theorematis fundamentalis in doctrina de residuis quadraticis 
demonstrationes et ampliationes novae, 
Comment. Soc. Reg. Sci. Gottingensis 4, 1816-1818.
Presented to the society Feb. 10, 1817.
[German translation:
Neue Beweise und Erweiterungen des Fundamentalsatzes in der
Lehre von den quadratischen Resten, in C.F. Gauss,
Untersuchen \"uber H\"ohere Arithmetik, ed. H. Maser,
New York: Chelsea, 1965, pp. 496-510.]

\item{\HK.}
F. Halter-Koch,
Quadratic Irrationals: An Introduction to Classical Number Theory,
CRC Press, 2013.

\item{\IR.}
K. Ireland and M. Rosen,
A Classical Introduction to Modern Number Theory,
Springer-Verlag, 1982. 

\item{\JACV.}
C. G. J. Jacobi,
Vorlesungen \"uber Zahlentheorie (Wintersemester 1836/37, K\"onigsberg),
ed. F. Lemmermeyer and H. Pieper, Erwin Rauner Verlag, Augsburg, 2007.

\item{\JACU.}
C. G. J. Jacobi,
\"Uber die Kriestheilung und ihre Anwendung auf die Zahlentheorie,
Bericht über die zur Bekanntmachung geeigneten Verhandlungen der 
Königl. Preuss. Akademie der Wissenschaften zu Berlin,
pp. 127-136, 1837.  Reprinted with comments and numerical tables
in J. Reine angew. Math., v. 30, 166-182, 1846.

\item{\KR.}
E. Kaltofen and H. Rolletschek,
Computing greatest common divisors and factorizations in quadratic
number fields, Math. Comp. 53, 697-720, 1989.

\item{\LEH.}
D. H. Lehmer, Guide to Tables in the Theory of Numbers, 
National Academy of Sciences, Washington, 1941.

\item{\LEM.}
F. Lemmermeyer, Reciprocity Laws: From Euler to Eisenstein,
Springer-Verlag, 2000,

\item{\MULL.}
S. M\"uller,
On the computation of cube roots mod $p$,
in A. van der Poorten, A. Stein, eds., 
High Primes and Misdemeanours: Lectures in Honour of the 60th Birthday 
of Hugh Cowie Williams (Fields Institute Communications v. 41), AMS,
2004.

\item{\SW.}
R. Scheidler and H. C. Williams,
A public-key cryptosystem utilizing cyclotomic fields,
Designs, Codes, and Cryptography 6, 117-131 (1995).

\item{\SHA.}
J. O. Shallit,
On the worst case of three algorithms for computing the Jacobi symbol,
J. Symbol. Comput. 10, 593-610 (1990).

\item{\SMIT.}
H. J. S. Smith, Report on the Theory of Numbers, Part I,
Report of the British Association for 1859, pp. 228-267.
Parts I-VI reprinted as H. J. S. Smith, Report on the 
Theory of Numbers, Chelsea, New York, 1965.

\item{\WEIL.}
A. Weilert, Fast computation of the biquadratic residue symbol,
J. Number Theory 96, 133-151, 1992.

\item{\WIKS.}
D. Wikstr\"om, On the $\ell$-ary GCD algorithm and computing
residue symbols, KTH (Stockholm) report TRITA NA 04-39, May 2004.


\item{\WILL.}
H. C. Williams, An $M^3$ public-key encryption scheme,
Proc. CRYPTO 85, LNCS 218, pp. 358-368 (1986).

\item{\WH.}
H. C. Williams and R. C. Holte, Computation of the solution
of $x^3 + D y^3 = 1$, Math. Comp. 31, 778-785 (1977).

\item{\WIN.}
S. Winograd, Arithmetic Complexity of Computations,
SIAM, 1980.


\vfill
\eject

\end

\obeylines

C. G. J. Jacobi,
Vorlesungen \"uber Zahlentheorie (Wintersemester 1836/37, K\"onigsberg),
ed. F. Lemmermeyer and H. Pieper, Erwin Rauner Verlag, Augsburg, 2007.

Bach-Shallit Chap. 3 gives a complexity bound to multiply n1 and 
n2 bit numbers: Let M = max, m = min of these.  Then it is 
O(M log m loglog m ) bops
Division is the same, but with M = max of quotient, divisor length,
m = min of these.

\end